\documentclass[9pt,sigconf,nonacm,screen]{acmart}

\usepackage[all]{nowidow}
\usepackage{xcolor}
\usepackage{subcaption}
\usepackage{hyperref}
\usepackage{acronym}

\usepackage[capitalise]{cleveref}
\crefformat{section}{\S#2#1#3} 
\crefformat{subsection}{\S#2#1#3}
\crefformat{subsubsection}{\S#2#1#3}

\begin{document}

\author{Marvin Kruber}
\affiliation{%
    \institution{TU Berlin \& ECDF}
    \city{Berlin}
    \country{Germany}}
\email{mkr@mcc.tu-berlin.de}
\orcid{0009-0005-6026-5718}

\author{Tobias Pfandzelter}
\affiliation{%
    \institution{TU Berlin \& ECDF}
    \city{Berlin}
    \country{Germany}}
\email{tp@mcc.tu-berlin.de}
\orcid{0000-0002-7868-8613}

\author{David Bermbach}
\affiliation{%
    \institution{TU Berlin \& ECDF}
    \city{Berlin}
    \country{Germany}}
\email{db@mcc.tu-berlin.de}
\orcid{0000-0002-7524-3256}

\title{A Hybrid Communication Approach for Metadata Exchange in Geo-Distributed Fog Environments}

\begin{abstract}
    Metadata exchange is crucial for efficient geo-distributed fog computing.
    Existing solutions for metadata exchange overlook geo-awareness or lack adequate failure tolerance.
    We propose HFCS, a novel hybrid communication system that combines hierarchical and peer-to-peer elements, along with edge pools.
    HFCS utilizes a gossip protocol for dynamic metadata exchange.

    In simulation, we investigate the impact of node density and edge pool size on HFCS performance.
    We observe a performance improvement for clustered node distributions, aligning well with real-world scenarios.
    HFCS outperforms a hierarchical and a P2P approach in task fulfillment at a slight cost to failure detection.
\end{abstract}

\maketitle

\section{Introduction}
\label{sec:introduction}

Fog computing combines the infinite computation and storage resources of the cloud with the benefit of low access latency of edge nodes.
However, edge nodes are severely limited regarding their resources and prone to unavailability due to physical impacts~\cite{paper_bermbach2017_fog_vision,bonomi2012fog}.
For this reason, a fog system should distribute tasks in consideration of the underlying network and hardware conditions, i.e., the available capacity of a node and its physical proximity to client~\cite{paper_pfandzelter2021_zero2fog,pfandzelter2023fred}.

Consequently, the nodes have to exchange their metadata information.
Metadata can be \emph{static} or \emph{dynamic}:
Static metadata relates to properties that do not vary at runtime, such as the geographical location.
In contrast, dynamic metadata includes attributes which change at runtime, e.g., available computing power, bandwidth, latency to other nodes, or storage capacity~\cite{poster_pfandzelter2022_coordination_middleware}.
Dynamic metadata thus requires a continuous exchange.
There is a multiplicity of approaches to handle the challenge of the communication of metadata information~\cite{chekired2018,kademlia,santos2018,skarlat2018,poster_pfandzelter2022_coordination_middleware}, yet most of them ignore information about the geographical position of nodes, which is an important aspect of the efficiency in geo-distributed systems~\cite{paper_hasenburg2020_disgb,paper_hasenburg2019_geocontext}.
Existing solutions often also establish a single point of failure where large parts of the system are unavailable in case of a node failure~\cite{communicationInTheFog}.

In this paper, we present \emph{Hybrid Fog Communication System (HFCS)}, an efficient, scalable, and reliable communication mechanism for metadata information in geo-distributed fog computing systems~(\cref{sec:architecture}).
We implement a prototype of HFCS and examine how different parameters influence its performance.
We further compare HFCS with various communication approaches~(\cref{sec:evaluation}).

\section{Architecture of HFCS}
\label{sec:architecture}

\begin{figure}
    \centering
    \includegraphics[width=\linewidth]{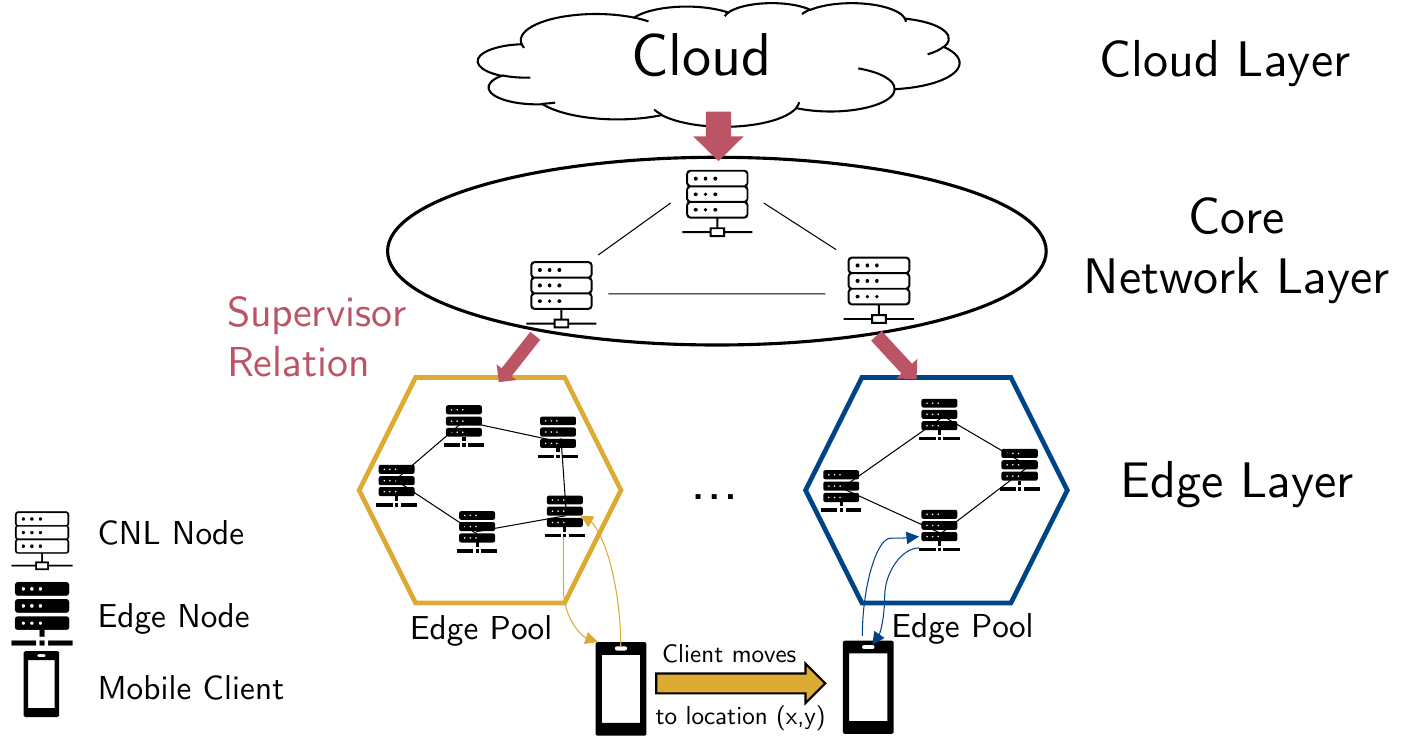}
    \caption{Architecture of HFCS}
    \label{fig:hfcs}
\end{figure}

As shown in \cref{fig:hfcs}, the architecture of HFCS is similar to the infrastructure of fog computing environments.
It comprises a hierarchy of cloud, core network, and edge layers.

\subsubsection*{Edge Layer}
\label{sec:edge_layer}

Multiple small peer-to-peer environments in the edge layer (\emph{edge pools}) keep the network overhead and transmission latency for nodes in proximity low while optimally distributing tasks.
An edge pool is a set of edge nodes grouped together to form a network based on their geographical proximity to each other.
If the capacity of the node a client connects to would be exceeded by an arriving task, the task can be redirected to another node in the edge pool.
If a client exits the geofence of the current edge pool, this procedure is escalated to the core network layer, where the closest edge pool with the appropriate edge node is determined.

\begin{figure}
    \centering
    \includegraphics[width=\linewidth]{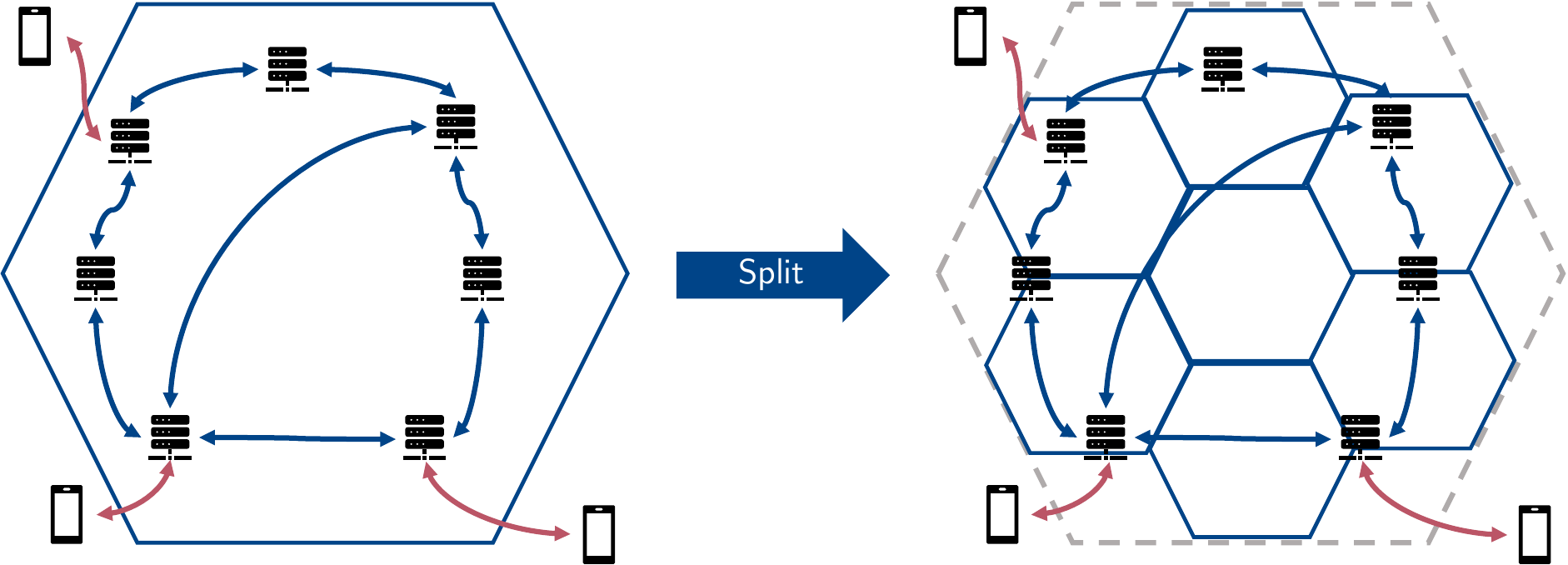}
    \caption{Split of an edge pool: if the maximum number of edge pool members would be exceeded by inserting a new node, the edge pool is divided into seven subpools with individual nodes assigned to their closest subpool.}
    \label{fig:split}
\end{figure}

We adopt the \emph{H3} grid system~\cite{h3} for edge pool structure by using a hexagonal geofence aligned to form a global grid system.
The hexagonal shape of edge pools simplifies the calculation and analysis as it keeps the distance between two pools (from center to center) equal to all neighbors.
Edge pools have a maximum number of members to achieve a balance between metadata information propagation within the pool and network overhead.
If this size is exceeded, the edge pool is subdivided into seven hexagonal sub pools (\cref{fig:split}) to overcome the challenge of geographical lumps of edge nodes, which is particularly relevant in urban areas.

Edge nodes only communicate with members of the same pool.
They exchange metadata information in fixed time intervals by using a gossip protocol~\cite{kermarrec2007gossiping}.
We implement a version vector~\cite{lamportClock} for each dynamic metadata information field of a node to ensure synchronization and causality conservation.

\subsubsection*{Core Network Layer}
\label{sec:core_network_layer}

Each edge pool is supervised by a core network layer node (\emph{CNL node}) responsible for the edge pool life cycle.
CNL nodes are also organized in a peer-to-peer manner but without size limitation.
Analogously to the edge layer, CNL nodes employ a gossip protocol for versioned metadata information exchange.

\subsubsection*{Cloud Layer}
\label{sec:cloud_layer}

At the root of the hierarchy, the cloud supervises CNL node life cycle, i.e., insertion, runtime management, and deletion.
It also determines the closest CNL node for new clients and instructs to find the corresponding closest edge node.
If a suitable edge node exists, its address is returned to allow the mobile client to start regular communication.
Otherwise, the address of the CNL node itself is sent to the client.
The cloud also accepts requests for the registration of new edge nodes and passes them to the respective closest CNL node in the form of an insertion order.

\subsubsection*{Failure Handling}
\label{sec:corner_cases}

Node failures for edge nodes can be detected if gossip partners do not receive answers or updates from a node within a specified time frame.
We implement a supervisor escalation mechanism for failure handling where the supervisor is contacted in case of a missing reply.
If a supervisor receives at least two messages of different subordinated nodes within a specific time range, it starts a failure routine for the failed node.

If a CNL node fails, all supervised edge pools would be separated from the system.
To avoid such a single point of failure, CNL nodes extend metadata exchange messages with their supervised edge pools.
An edge pool thus stays a part of the system, as every previous gossip partner of the failed CNL node is able to become the new supervisor.
The final assignment is then done by the cloud.

In contrast to the cloud, fog and edge nodes have limited capacity.
If a client contacts a node with insufficient capacity to complete its task, it either redirects the task to other nodes of the same subnetwork or escalates it to the supervisor, which has more resources.
Suppose there is a fixed-size edge pool and the requested node is fully occupied.
Due to the regular metadata information exchange the edge node knows the capacity of the other nodes within the pool and can determine a node with the required resources and requests to forward the task to it.
If the target node accepts, the task is redirected and the latency increases minimally through the geographical closeness of the nodes.
Otherwise, the initial node sends a request to another node.
If this procedure does not yield a solution, the task can be escalated.

\section{Evaluation}
\label{sec:evaluation}

We implement a Java prototype of HFCS to evaluate our approach\footnote{\url{https://github.com/OpenFogStack/HFCS}} (\cref{sec:simulation}).
We examine how the node density  (\cref{sec:density}) and edge pool size (\cref{sec:pool_size}) affects the message volume and the failure tolerance of HFCS, how the edge pool size.
We then compare HFCS to alternative node communication approaches, namely hierarchical, tree-like communication and regular broadcasts (\cref{sec:comparison}).

\subsection{Simulation Scenario and Prototype}
\label{sec:simulation}

Our HFCS prototype uses a \emph{gossip thread}, which handles the metadata information propagation of a node, and \emph{worker threads} that simulate task processing with different hardware capabilities at different layers of the fog continuum.
The gossip thread implements a gossip protocol, with initiations per node limited to three and a gossip time interval of three seconds.
Client threads send tasks at intervals of one to five seconds and then simulate random movement to a new position by a random distance of up to two geographical longitudes and/or latitudes.
Further, we inject failures among edge (90\%) and CNL nodes (10\%) every 50 to 500 seconds.

We perform five measurement iterations per value.
Each measurement iteration lasts 15 minutes (plus 10 minutes for initialization and data storage activities).
There are 1500 clients, 1000 edge nodes, 50 CNL nodes and one cloud, which are distributed randomly around the world.
Based on the geographical coordinates, we assign the nodes to continents, which are assumed as rectangles for simplicity, with different probabilities.

\subsection{Node Density}
\label{sec:density}

In the first experiment, we examine the performance of HFCS for various node densities.
For this, we set the maximum distance \emph{d} between two nodes as $0.5$, $2$, $5$, $10$, and $20$.

\subsubsection*{Message Volume}

\begin{figure}
    \centering
    \includegraphics[width=\linewidth]{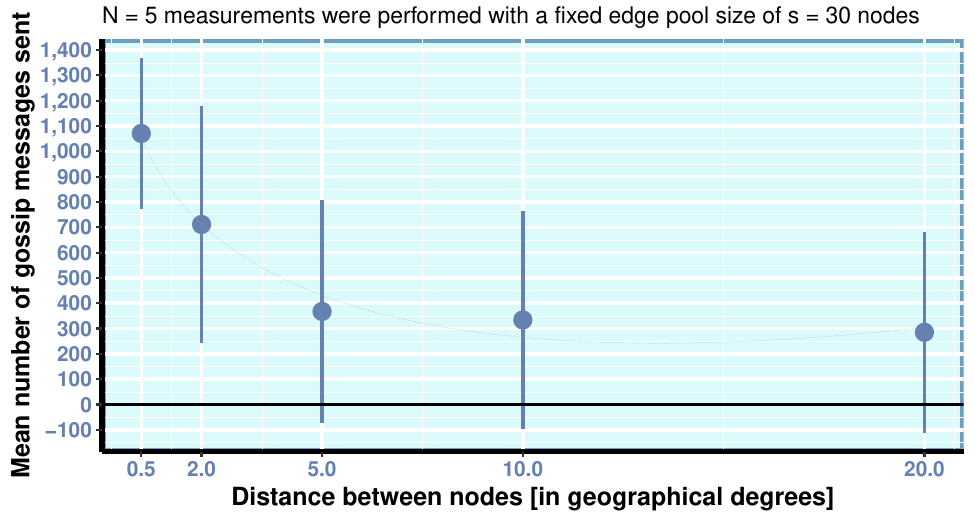}
    \caption{The average volume of exchanged gossip messages decreases exponentially with the node distance due to the more scattered distribution of nodes across the edge pools.
        It also leads to an increasing standard deviation up to $d = 10$.}
    \label{fig:density_gossip}
\end{figure}

\begin{figure}
    \centering
    \includegraphics[width=0.8\linewidth]{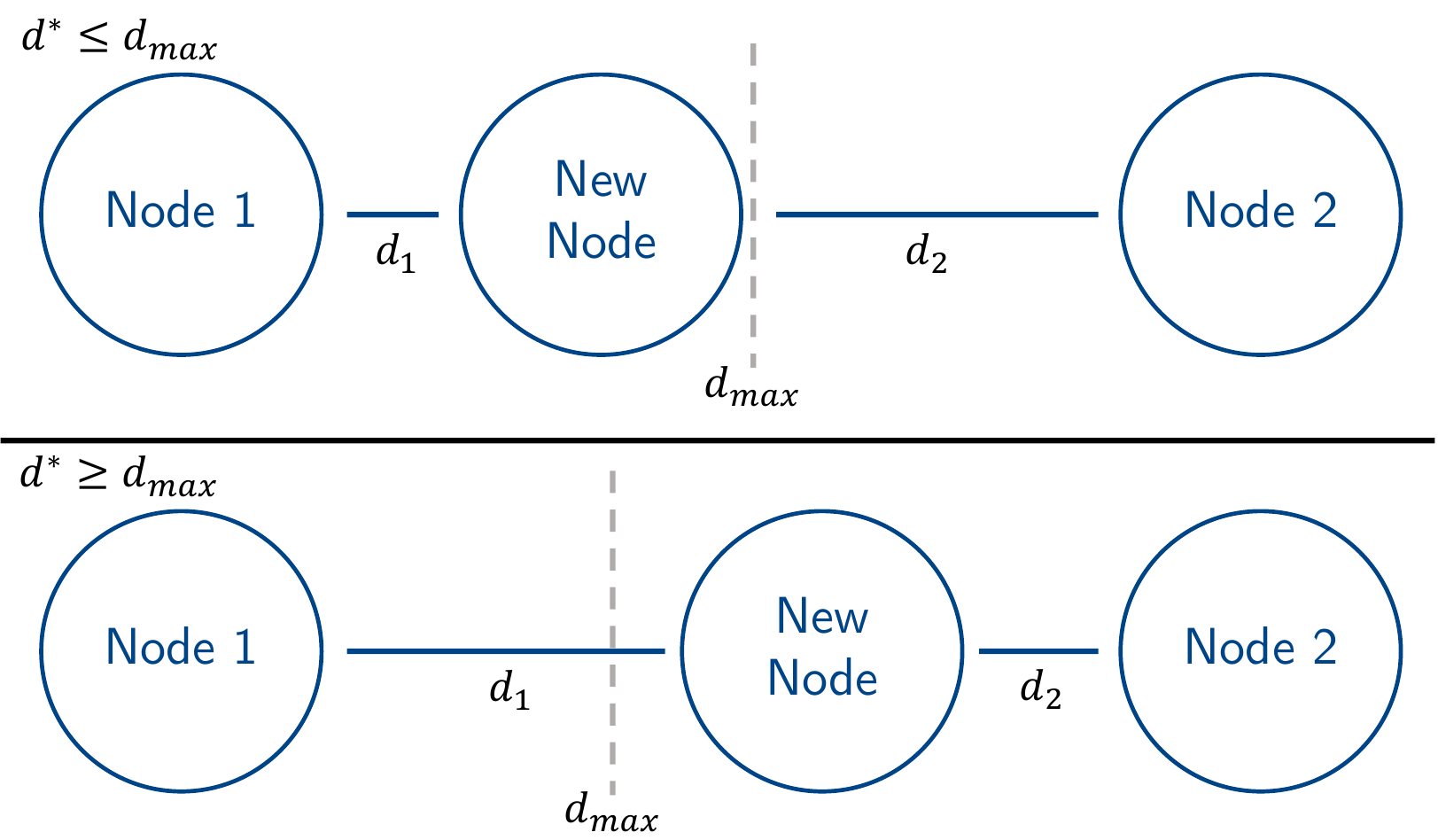}
    \caption{With continental restrictions there is a distance $d_{max}$ for which $d_1 = d_2$.
        For any distance between nodes $d > d_{max}$, the maximization of
        the distance $d_1$ would simultaneously lead to a minimization of $d_2$.}
    \label{fig:distance_dilemma}
\end{figure}

\begin{figure*}
    \begin{subfigure}{0.5\linewidth}
        \centering
        \includegraphics[width=\linewidth]{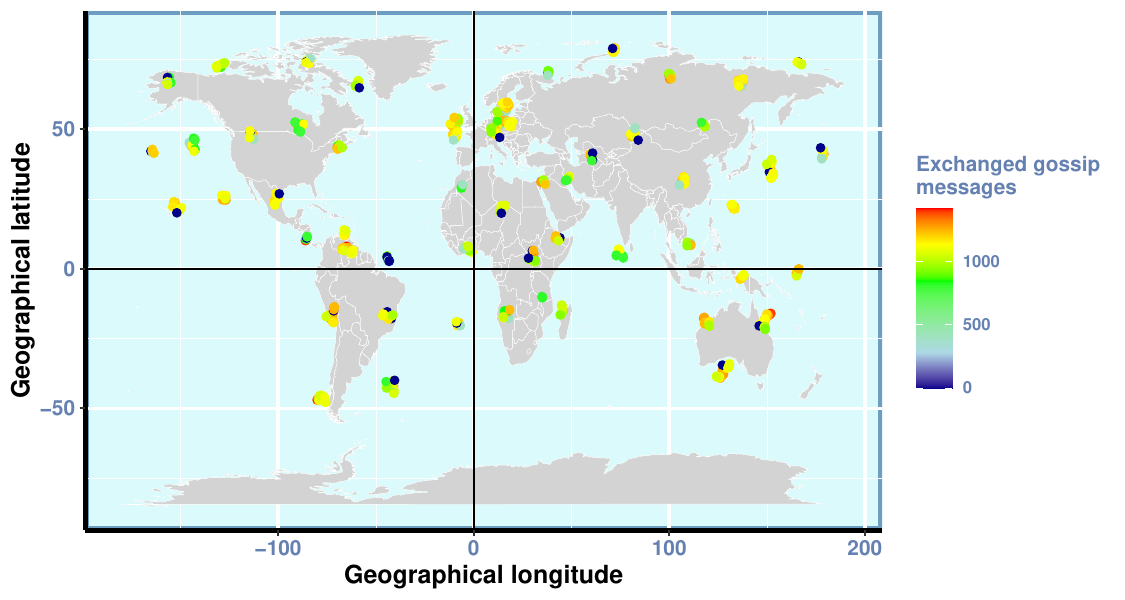}
        \subcaption{$d = 0.5$}
    \end{subfigure}%
    \hfill
    \begin{subfigure}{0.5\linewidth}
        \centering
        \includegraphics[width=\linewidth]{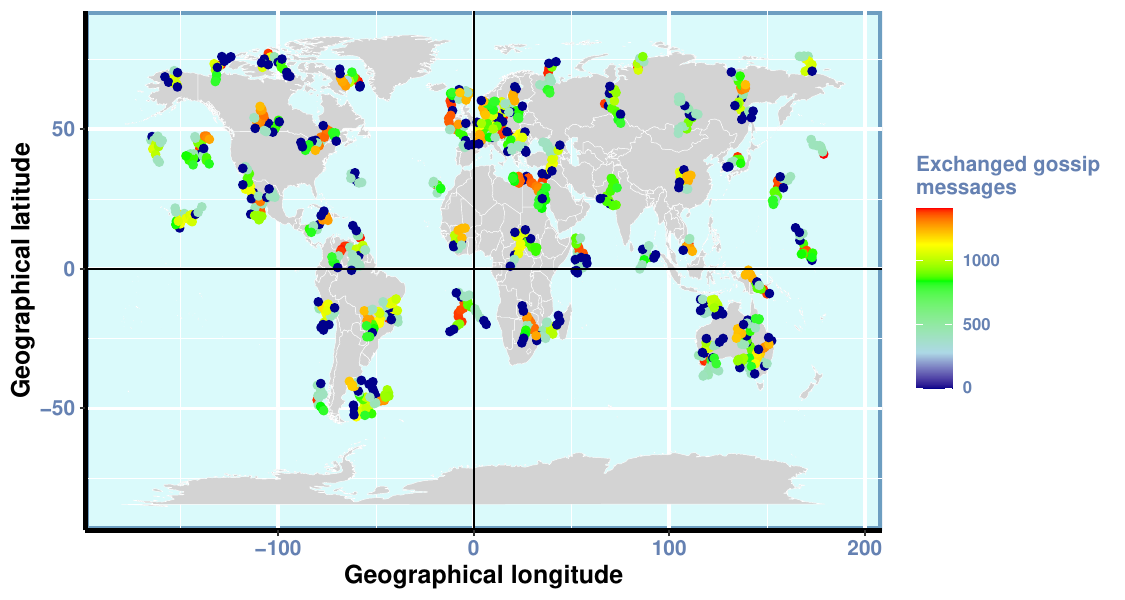}
        \subcaption{$d = 2$}
    \end{subfigure}%
    \vfill
    \begin{subfigure}{0.5\linewidth}
        \centering
        \includegraphics[width=\linewidth]{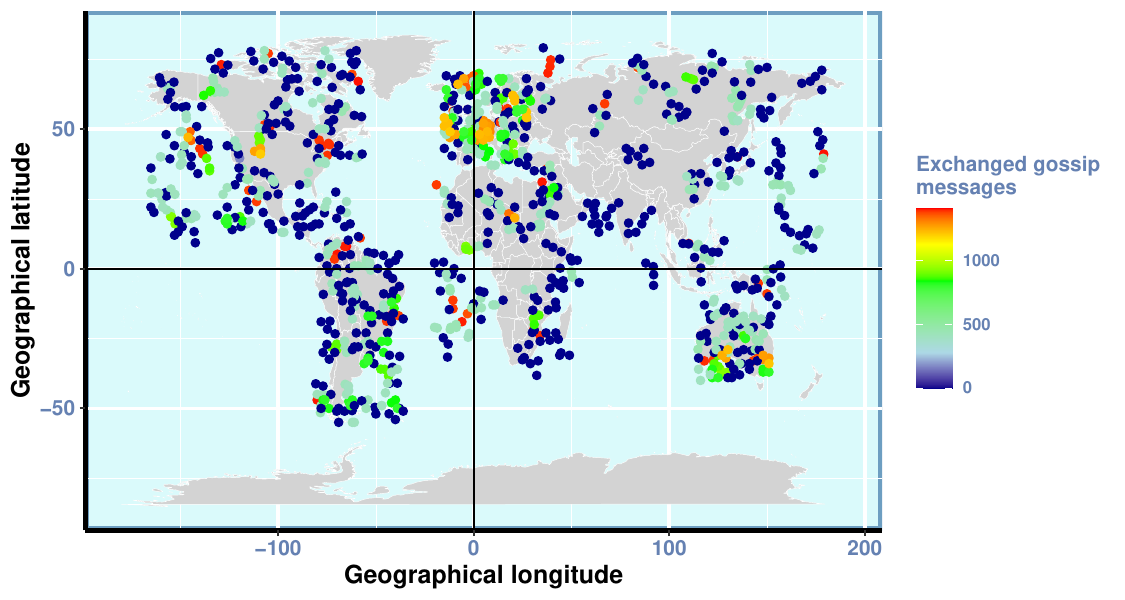}
        \subcaption{$d = 5$}
    \end{subfigure}%
    \hfill
    \begin{subfigure}{0.5\linewidth}
        \centering
        \includegraphics[width=\linewidth]{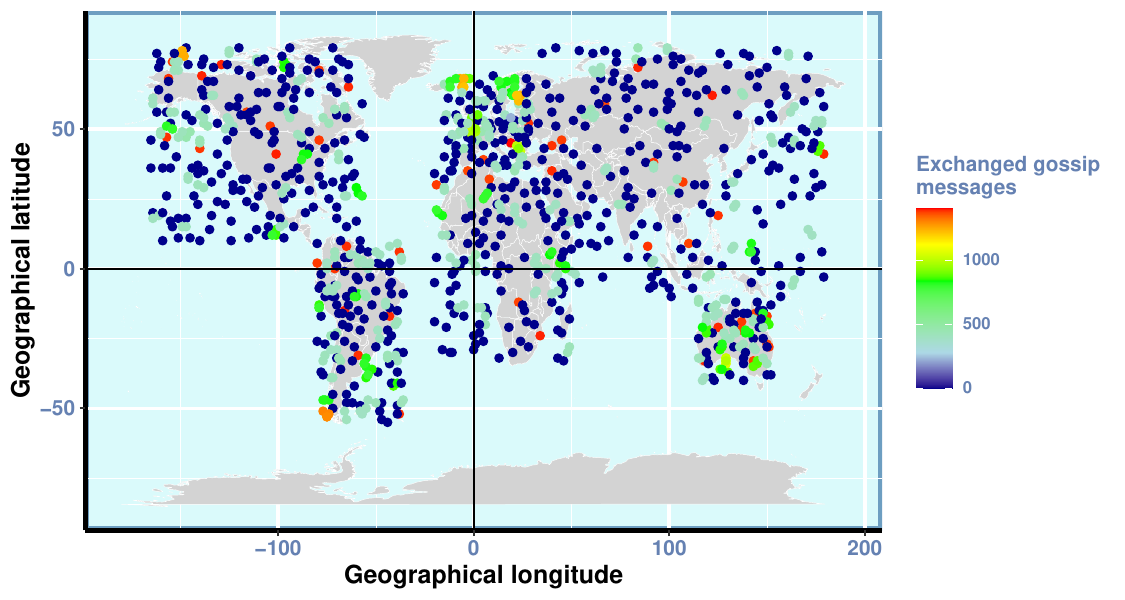}
        \subcaption{$d = 20$}
    \end{subfigure}%
    \caption{Geographical distribution of the message exchange volumes of nodes under various maximum node distances $d$ considering agglomerations and continental restrictions with continents assumed as rectangles.
        Low message exchange volumes of individuals nodes in agglomerations are caused by the split of edge pools into seven subpools which affects either an isolation of individual nodes, or a reduction of communication partners in the corresponding pool so that their message volume in total grows more slowly.
        Large maximum distances between nodes lead to a more scattered distribution of edge nodes to various pools which also causes a slower growth of the message exchange volume.}
    \label{fig:distribution_metric}
\end{figure*}

\cref{fig:density_gossip} shows the relation between the maximum distance between nodes and the mean number (combined with the standard deviation) of gossip messages a node has exchanged.
The message exchange volume per node decreases exponentially with an increasing node distance.
This correlation can be traced back to the conception of the edge pools.
An edge pool contains all edge nodes which are located in a certain vicinity.
The increasing distance between the nodes causes a new distribution of the edge nodes to various edge pools, so that the number of non-empty pools grows whereas the average number of members of non-empty pools diminishes.
This leads to lower communication overhead but also to a lower metadata information diffusion inside the whole system as well as a reduction of possible redirection options a node can use in case of overload.
This also explains the rising standard deviation up to a node distance of 10, as there is a more varying number of members in non-empty pools.
For a maximum node distance of $20$, the standard deviation of exchanged gossip messages per node seems to lower.
Responsible for this effect are our continental distribution assumptions:
If nodes only can be placed on a continent, we implicitly assume an area boundary.
Then, the maximum possible distance between nodes (provided a distribution of $n$ nodes on this continent) is limited as well.
For any distance value which is larger than this limit, the maximization of the distance to the closest node would lead simultaneously to the minimization of the space to another edge node, as shown in~\cref{fig:distance_dilemma}.

We show the concrete distribution of message volumes of the nodes for the maximum node distance of $0.5$, $2$, $5$, and $20$ in \cref{fig:distribution_metric}.
In high node density scenarios, there are nodes with a low number of exchanged messages in agglomerations.
We trace this back to the concept of splitting an edge pool into seven subpools if the maximum number of pool members are exceeded.
Hence, it is possible that some nodes are either isolated in a subpool or have few communication partners in their pool so that their message volume does not grow as much compared to nodes in other subpools.
For large maximum node distances, low message exchange volumes of nodes are reasoned by the more scattered distribution of edge nodes to various edge pools, as mentioned earlier.

\subsubsection*{Task Completion}

\begin{figure}
    \centering
    \includegraphics[width=\linewidth]{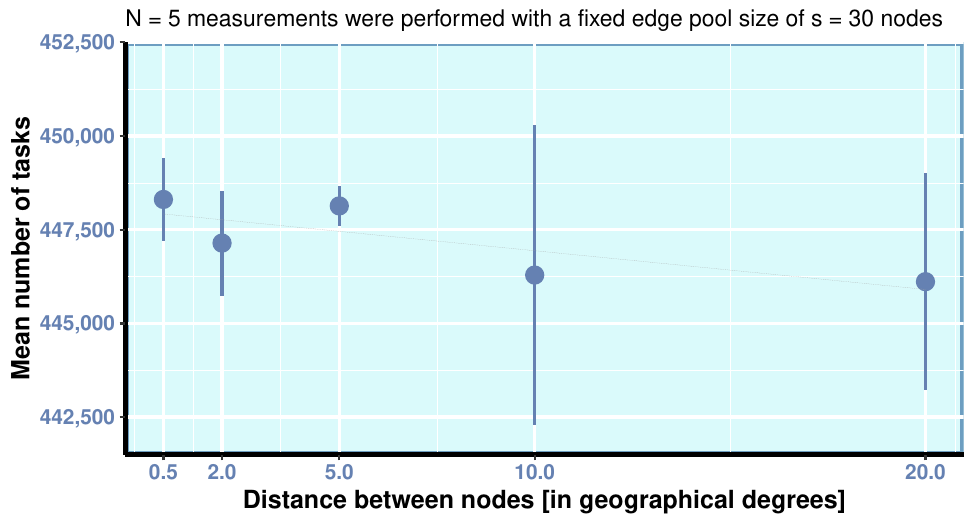}
    \caption{The average number of completed tasks decreases with larger distance between nodes.
        The anomaly for $d = 5$ can be traced back to the non-deterministic
        behavior of threads.
        The irregularity for $d = 20$ is a consequence of the continental distribution restrictions.}
    \label{fig:density_completed}
\end{figure}

We show the influence of the node density on the mean total number of completed tasks in \cref{fig:density_completed}.
More tasks are completed as maximum distance between two nodes increases, as larger maximum node distance leads to larger pools with higher total capacity.
Redirected tasks are thus also more likely to be completed after the first redirection.
Similarly, the standard deviation also rises with the maximum node distance.
We attribute the anomaly at $d = 5$ to the non-deterministic behavior of threads.

\begin{figure}
    \centering
    \includegraphics[width=\linewidth]{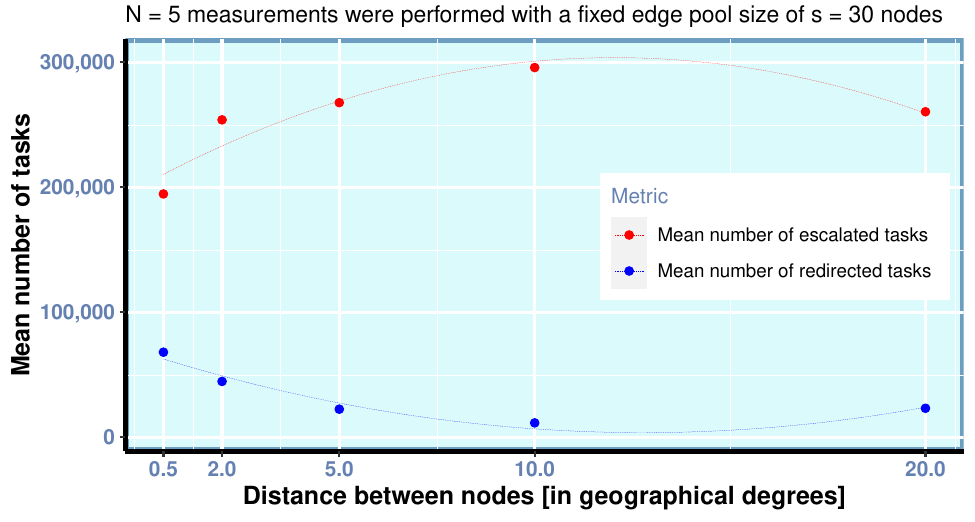}
    \caption{The mean number of redirected tasks decreases with growing distance between nodes while the number of escalated tasks increases. The parabola courses are reasoned by the reduction of possible redirection options for a more scattered distribution of nodes.
        Hence, more tasks are escalated if a node is overloaded.
        The reversal of this effect for $d = 20$ is a consequence of the continental distribution.}
    \label{fig:density_mean_redirected}
\end{figure}

As depicted in \cref{fig:density_mean_redirected}, the mean number of escalated tasks is always greater than the average number of redirected tasks, i.e., in each tested scenario the shared capacities of an edge pool are mainly sufficient.
The more tasks are escalated, the fewer tasks are redirected and vice versa.
The frequency of task escalation compared to redirection increases with a growing maximum distance between the nodes.
It is caused by the reduction of possible redirection options within an edge pool due to the more scattered distribution of nodes.
The anomaly for $d = 20$ is again a consequence of the spatial restrictions of the distribution of nodes.
The corresponding standard deviations tend to rise with a growing maximum space between nodes.
An irregularity can only be observed for $d = 5$, where the mean number of escalated tasks increases but simultaneously the associated standard deviation decreases.
This phenomenon persisted in control measurements and is likely a side effect of the standard edge pool distance of 5 geographical units.

\subsubsection*{Failure Detection}

\begin{figure}
    \centering
    \includegraphics[width=\linewidth]{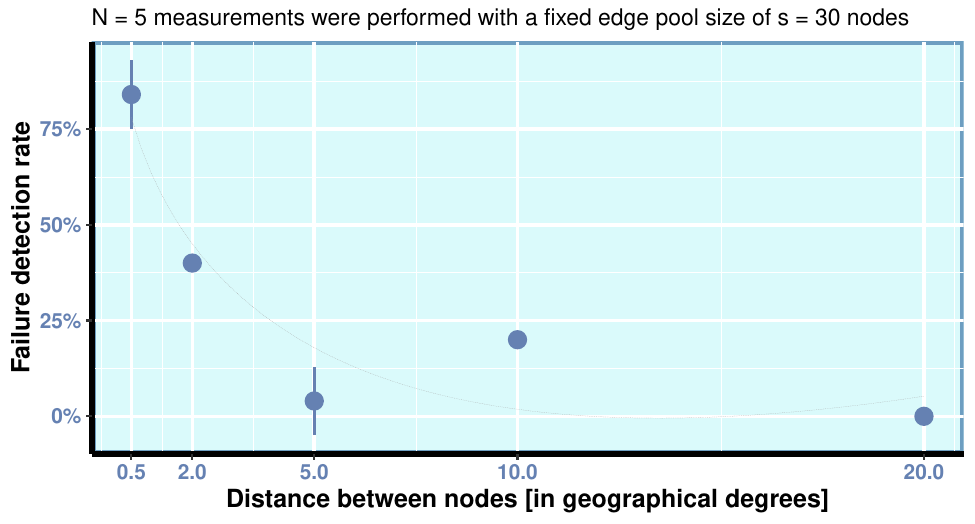}
    \caption{The failure detection rate decreases exponentially with a more scattered distribution of nodes because of the consensus mechanism for failure identification in edge pools.
        Deviations can be traced back to the non-deterministic behavior of threads and a comparatively slow failure detection.}
    \label{fig:density_failure}
\end{figure}

\cref{fig:density_failure} show the relation between the failure detection rate and the node density.
The mean failure detection rate decreases exponentially with a growing distance between nodes, likely caused by the failure detection mechanism (\cref{sec:corner_cases}).
As two nodes in an edge pool must both report node failure to the supervisor, there must be at least three edge nodes (including the failed one) in a pool for a successful detection.
This also explains why the maximum failure detection amounts to 84\%.
We derive that the node density has a direct influence on the system's performance due to the concept of edge pools.
HFCS synergizes well with agglomerated distributions of edge nodes, which tends to correspond to reality.

\subsection{Pool Size}
\label{sec:pool_size}

Another parameter which we considered in our experiments is the edge pool size, i.e., the maximum number of nodes contained as well as the geographical range of the pool.
Based on the results of \cref{sec:density} and in order to model the case of agglomerations appropriately, we choose $d = 0.5$.
We simulate scenarios in which the edge pools can comprise up to $10$, $20$, $30$, $40$, and unlimited members.
Additionally, we tested a scenario where all edge nodes are grouped together into a single, global edge pool.

We observe that the average number of exchanged gossip messages per node such as the standard deviation are not affected by an increment of the limit of pool members.
However, the global instance exhibits a significantly higher mean volume of exchanged gossip messages per node and a lower standard deviation.
This can be explained by a more even distribution of message exchanges, which results from the integration of all nodes into the global pool.

\begin{figure}
    \centering
    \includegraphics[width=\linewidth]{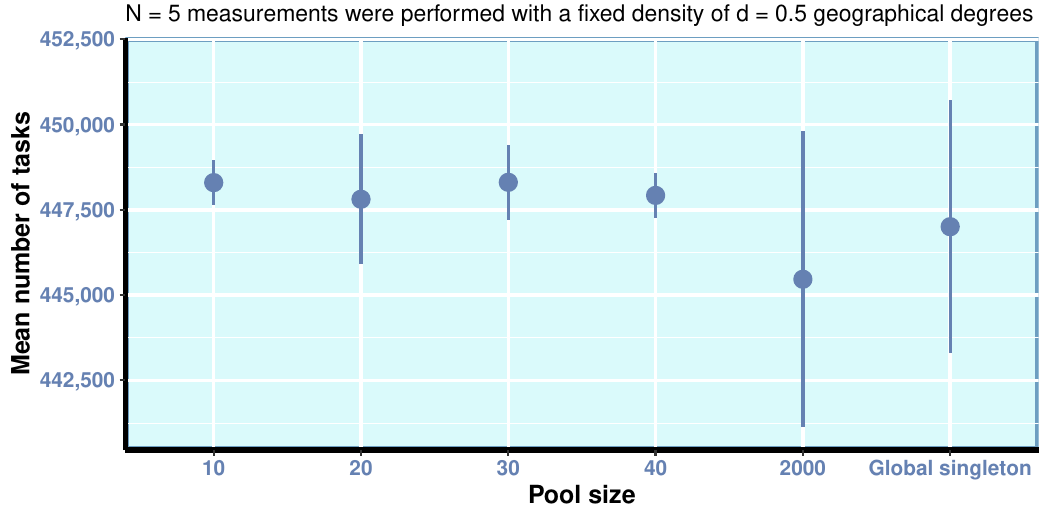}
    \caption{
        The pool size of $s = 30$ members achieves the best performance on average.
        An increasing number of edge pool participants leads to a rising degree of information inefficiency due to a slower metadata information diffusion in the pool and by that an increasing number of unsuccessful redirection requests.
        The global edge pool instance performs better than unlimited edge pools owing to the additional capacity gains through the integration of all edge nodes.
    }
    \label{fig:eps_completed}
\end{figure}

\cref{fig:eps_completed} illustrates the mean number of completed tasks in dependency of the edge pool size.
The maximum value of $448,307$ mean completed tasks is reached with a limit of $30$ members.
Moreover, the standard deviation tends to decrease with a growing ceiling of pool members.
Performance decreases with unlimited number of pool members and the global edge pool instance given the trade-off between a fast metadata information diffusion within the pool and the communication costs of gossip.

\begin{figure}
    \centering
    \includegraphics[width=\linewidth]{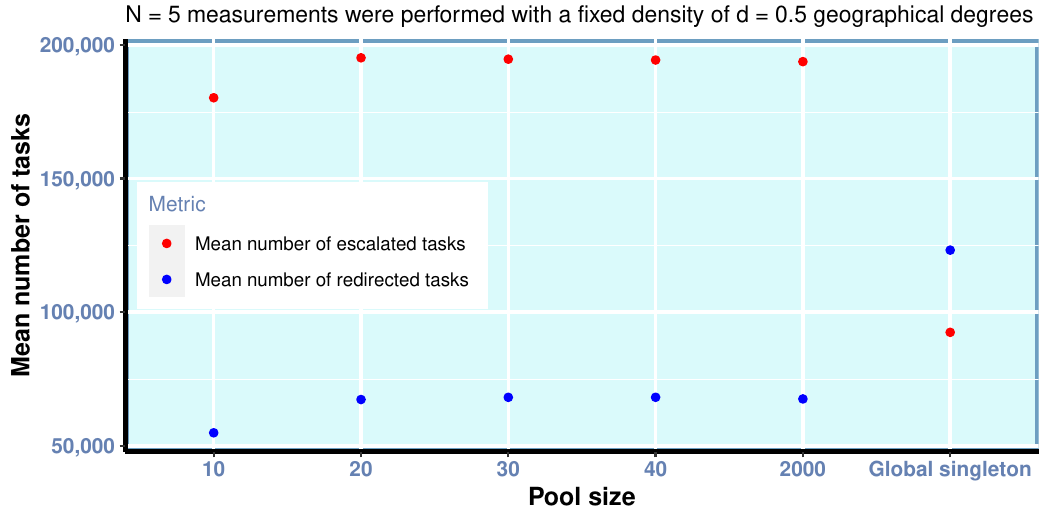}
    \caption{The average numbers of redirected, and escalated tasks are generally unaffected by the pool size.
        A more equal distribution of the task load is achieved for $s = 10$ due to finer granular subpools and by that more frequent pool-hopping of clients.
        The global edge pool instance redirects and escalates fewer tasks on average by virtue of the larger capacity contingent through the integration of all edge nodes.}
    \label{fig:eps_mean_redirected}
\end{figure}

As shown in \cref{fig:eps_mean_redirected}, the mean number of redirected and escalated tasks are generally unaffected by an increasing number of possible pool members.
At a maximum edge pool size of $10$, the mean numbers of redirected and escalated tasks are lower, while the number of completed tasks is almost as high as the maximum.
While appearing contradictory, pools tend to split more frequently for a limited pool size, provided the density is constant.
Clients thus change edge pools more frequently, leading in total to a more evenly distributed task load within the whole edge layer, and fewer overloaded capacity of edge pools or subpools.
Further, the global edge pool instance has significantly fewer redirected and escalated tasks on average as a result of its larger total capacity.

In summary, the experiment reveals that the number of edge pool members does not influence the number of exchanged gossip messages.
However, the replacement of edge pools by a single global peer-to-peer network within the edge layer seems to have a negative impact on the volume of exchanged gossip messages.
The failure detection rate is unaffected by the pool size.

\subsection{Comparison}
\label{sec:comparison}

Finally, we compare HFCS with a maximum node distance of $d = 0.5$ and upper bound of $30$ nodes per pool with a \emph{hierarchical system}, where child nodes communicate in regular time intervals with their parent node and send their metadata information to it, and a \emph{peer-to-peer} (P2P) system using broadcasting among all nodes.

\begin{figure}
    \centering
    \includegraphics[width=\linewidth]{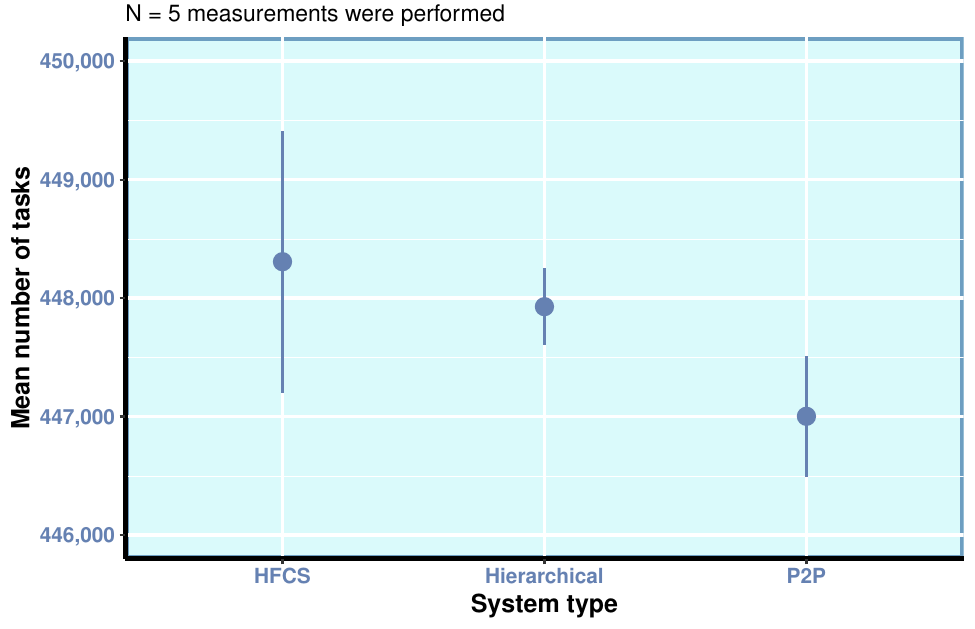}
    \caption{HFCS achieves the best performance but also has the highest standard deviation.
        The communication expense of the P2P approach limits its task fulfillment.}
    \label{fig:b_completed}
\end{figure}

The hierarchical system requires a mean $437$ exchanged gossip messages per node, followed by HFCS with $1,070$ messages, and the P2P system with $442,652$ messages.
Trivially, the broadcast leads to the highest communication expense, but similarly it guarantees the highest and fastest metadata information diffusion on a system-wide level.
The broadcast in edge pools in HFCS requires more messages than strictly hierarchical message exchange.

\cref{fig:b_completed} illustrates the mean task completion of all three systems.
HFCS completes the most tasks on average but exhibits the highest standard deviation at the same time.
In contrast, the P2P approach performs significantly worse given the tremendous communication expense of the broadcast mechanism.

\begin{figure}
    \centering
    \includegraphics[width=\linewidth]{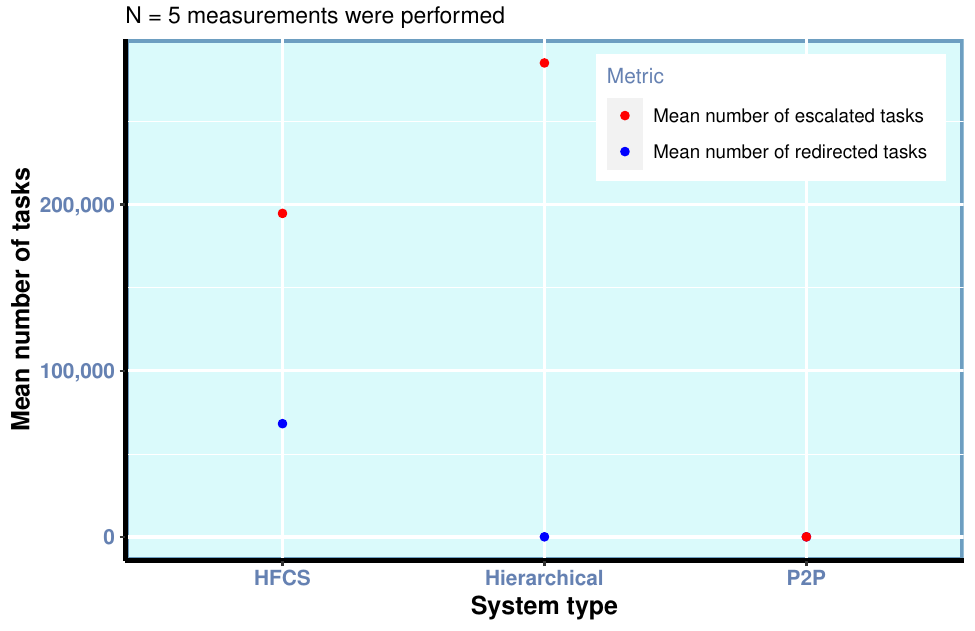}
    \caption{HFCS escalates fewer tasks than the hierarchical system but redirects more than the peer-to-peer prototype.
        There is no redirection of tasks in the peer-to-peer system         owing to an inconvenient combination of various factors, such as the equal ratio of clients to nodes, the request frequency of clients, the geographical distribution and the capacity increase caused by the integration of the cloud and the CNL nodes.}
    \label{fig:b_mean_redirected}
\end{figure}

For conceptual reasons, no values greater than zero could be observed for either the number of redirected tasks of the hierarchical benchmark or for the number of escalated tasks in case of the peer-to-peer system.
The performances of the three systems for the corresponding metric are depicted in \cref{fig:b_mean_redirected}.
The maximum mean number of escalated task is discernible for the hierarchical benchmark.
It attains a number of $285,272$ escalated tasks on average.
However, HFCS only escalates $194,692$ tasks and redirects $68,120$ tasks on average.
The sum of both these numbers is circa $20,000$ lower than the mean number of escalated tasks of the hierarchical benchmark.
This implies that HFCS could have a better scaling effect than the hierarchical approach by distributing the load on nodes of the same layer (i.e., redirect tasks) before it is escalated upwards.
The previous statistic about the mean number of completed tasks (\cref{fig:b_completed}) seems to confirm this relation.

Another extreme is the average number of redirected tasks of the peer-to-peer system.
Unexpectedly, it has a value of zero.
Responsible for this are the numerical proportions between nodes and clients as well as the geographical distribution of both, and the request frequency of the clients.
The capacity gain of the network through the direct integration of the cloud and the CNL nodes into the ``base'' network should be considered.
The interaction of all these factors causes this phenomenon.

Both the tree-like structure and the broadcast mechanism enable fast 100\% failure detection whereas HFCS reveals some issues in this context as discussed in \cref{sec:density}.
Nevertheless, HFCS is able to maintain most of the system functionalities in the event of a node failure, so that a sufficient level of failure tolerance is guaranteed.
In comparison, the hierarchical approach has a single point of failure design in the core network layer.
If one node fails, all associated edge nodes become unavailable for the whole system.

\section{Related Work}
\label{sec:related_work}

We distinguish between \emph{hierarchical}, \emph{peer-to-peer}, and \emph{hybrid} approaches to node communication in fog environments

\subsubsection*{Hierarchical Approaches}

Tree-like communication structures, where communication between two child nodes are only possible via their parent, are often preferred in fog computing given the layered fog infrastructure~\cite{communicationInTheFog}.
Chekire et al.~\cite{chekired2018} describe a multi-layer fog computing system in which the nodes primarily communicate via the transmission of jobs.
They assume an architecture that arranges the nodes in layers in the hierarchy according to their computing and storage capacities, where the ones with the lowest resources build the leaves of the tree.
Their idea is to increase the efficiency of the system by escalating tasks bottom up, i.e., if a job exceeds the available resources of a leaf node it is passed onto a higher layer.
Additionally, they develop a probabilistic model which depicts the likelihood that a given workload would exceed the capacities of the servers in a certain layer.
Based on this, they provide an algorithm for distributing the tasks to the single layers~\cite{chekired2018,communicationInTheFog}.
Skarlat et al.~\cite{skarlat2018} modify the common tree structure by introducing a fog controller -- a central component which is directly subordinated to the cloud.
It is mainly responsible for coordination and administration such as the distribution of workloads or the registration of new nodes.
Furthermore, they build hierarchical clusters of fog nodes, sensors, and actuators, so-called \emph{fog colonies}, i.e., the client, edge, and core network layer are combined in these units which can deploy individual services~\cite{communicationInTheFog,skarlat2018}.
The tree-like communication structure of hierarchical approaches combined with the aggregation and storage of metadata information on parent nodes leads to data centralism.
This achieves low communication costs and fast and efficient decision-making but establishes a single point of failure.

\subsubsection*{Peer-to-peer Approaches}

In contrast, unstructured overlay networks are distinguished by their peer-to-peer character where nodes coordinate themselves and communicate autonomously~\cite{distributedSystems}.
All nodes have equal rights and only a partial view of the network.
Exchanging metadata uses gossip protocols in which information is sent to a randomly selected neighbor node, e.g., at a fixed time interval~\cite{kermarrec2007gossiping}.
\emph{Kademlia}~\cite{kademlia} relies on asynchronous messaging and an ID-based routing algorithm.
Each node is assigned an 160-bit ID used to address the node itself and outgoing messages.
Accordingly, the recipients of a message can note the ID and prove its existence.
However, only IDs of a certain distance in the key space are stored (together with IP address and UDP port) by the receiver, where the distance between two keys is calculated by executing a XOR.
Messages are then forwarded to the node in the routing list with the closest ID to the recipients ID.
Santos et al.~\cite{santos2018} extend this with a distributed hash table in order to exchange metadata, store it, and enable a dynamic resource provision by this.
Peer-to-peer systems do not introduce the data centralism of hierarchical approaches, but introduce high communication costs and ignore potential underlying hardware heterogeneity of nodes.

\subsubsection*{Hybrid Approaches}

Hybrid approaches combine hierarchical elements with the peer-to-peer character.
For this, a hierarchy of layers is implemented and the concepts of equality between as well as autonomy of all nodes within a layer are introduced~\cite{communicationInTheFog}.
Lee et al.~\cite{lee2017} provide a framework for the optimization of the formation of networks in the context of fog environments with the objective to minimize computational latency when outsourcing jobs or workloads to neighbor nodes through a dynamic selection of neighbors.
For this purpose, they solve an online secretary problem resulting in the mentioned hybrid communication architecture~\cite{communicationInTheFog,lee2017}.
Chen et al.~\cite{chen2017} suggest \emph{HyFog}, a framework which enables clients to decide where jobs should be executed.
They differentiate between the local execution, the outsourcing to physically close computing nodes (or other clients), and the offloading to the cloud.
Consequently, they assume an architecture consisting of a local layer, which equals a combination of device and edge layer, a device-to-device layer, i.e., core network layer, and a cloud layer.
Chen et al.~\cite{chen2017} aim to minimize the execution costs of tasks solving the offloading of jobs as a minimum weight matching problem.

\section{Conclusion}
\label{sec:conclusion}

Existing solutions for metadata communication in geo-distributed fog systems ignore either geo-awareness or reveal single points of failure despite their peer-to-peer elements.
We introduced HFCS, a novel hybrid communication system, which is characterized by the concept of edge pools and its geo-awareness and uses a gossip protocol for exchanging metadata information.

In simulation, we examined the influence of the node density and the edge pool size on the performance of HFCS.
For more agglomerated distributions of nodes, we observed a clear performance increase, which synergizes well with real-world conditions.
Further, comparison of HFCS with a hierarchical system and a P2P broadcast approach has shown its superior performance.
In future work, we plan to examine the influence of various network topologies in the core network layer and edge pools, the use of movement prediction to limit edge pool hopping, network distance regression, consensus-based failure detection, and early task redirection.

\begin{acks}
    Supported by the \grantsponsor{DFG}{Deutsche Forschungsgemeinschaft (DFG, German Research Foundation)}{https://www.dfg.de/en/} -- \grantnum{DFG}{415899119}.
\end{acks}

\balance
\bibliographystyle{ACM-Reference-Format}
\bibliography{bibliography.bib}

\end{document}